\documentstyle[12pt,epsfig,amssymb]{article}

\begin{document}

\begin{center}

{\Large \bf Statistical properties of the estimator using covariance matrix}

\vspace{0.1in}

{\bf S.I. Alekhin}

\vspace{0.1in}
{\baselineskip=14pt Institute for High Energy Physics, 142284, Protvino, Russia}

\begin{abstract}
The statistical properties of estimator using covariance matrix 
for the account of point-to-point correlations due to systematic 
errors are analyzed. It is shown that the covariance matrix estimator
(CME) is consistent 
for the realistic cases (when systematic errors on the fitted parameters are 
not extremely large comparing with the statistical ones)
and its dispersion is always smaller, than 
the dispersion of the simplified $\chi^2$ estimator 
applied to the correlated data.
The CME bias is negligible for the realistic 
cases if the covariance matrix is calculated during the fit iteratively 
using the parameter estimator itself. Analytical formula for the covariance 
matrix inversion allows to perform fast and precise calculations 
even for very large data sets. All this allows for efficient use 
of the CME in the global fits. 
\end{abstract}
\end{center}
\newpage

\section*{INTRODUCTION}

 Modern particle physics development becomes more and more based 
on the analysis
of precise experimental data. This demands refining of all stages
of the data inference including the account of 
correlations due to systematic uncertainties 
which are often comparable or even larger than the statistical ones.
In particular this problem is important for the 
precise tests of Standard Model 
and determination of the parton distributions \cite{:2000nr,Catani:2000jh}.
Many authors for the sake of simplicity very often use
approaches which ignore point-to-point correlations due to systematic
errors, i.e. sum all errors in quadrature or drop systematics
at all. It is evident that if the systematic errors are important source of
the data uncertainty such approaches can lead to the 
distortion of the estimated errors on the fitted parameters.
At the same time the construction of estimators accounting for the 
correlations is not straightforward since
the competitive probabilistic model of data can be used in the analysis. 
Essentially two generic models are possible: One based on the frequentist  
treatment of systematic shifts and another one based on the Bayesian 
approach. This paper is concentrated on the analysis of statistical 
properties of the estimators within the Bayesian treatment of 
systematic errors. An introduction
into this scope given in Ref.~\cite{D'Agostini:1995fv} contains 
argumentation in favor of this approach. The only point 
that we would like in particular underline here is that 
the Bayesian treatment is the only constructive way in
the case of many sources of systematics when classical treatment which implies
introduction of additional parameter for every source of systematic errors
can cause great problem with the 
interpretation/representation of the function of the large number of arguments.

The natural way to account for point-to-point correlations due to 
systematic errors within Bayesian approach is to use
covariance matrix associated with systematic errors 
(see e.g. Refs.~\cite{Swartz:1994qz,Gates:1995rq}).
Meanwhile, there are concerns that the covariance matrix estimator (CME)  
can result in biased values of the parameters values and their dispersions (see 
Refs.~\cite{Seibert:1994sf,Michael:1994yj,D'Agostini:1994uj,Swartz:1996hc}). 
In this connection it worth to recall 
that the estimators accounting for the data correlations 
often exhibit poor statistical properties
regardless they use covariance matrix or not.
For example as it was shown in Ref.~\cite{Daniell:1984ea} 
the sample dispersion estimated from the 
the correlated Monte Carlo data sets can 
acquire the bias equal to the dispersion value itself\footnote{This 
effect is connected with the 
well known fact that the sample dispersion gives biased 
estimation of the studied distribution dispersion;
the correlations merely amplify this bias.}. 
At the same time the estimators would be unbiased if the 
covariance matrix is not evaluated from the 
measurements itself. Indeed, the unbiased estimator 
for the correlated Monte-Carlo data
was constructed in Ref.~\cite{Michael:1995sz}
using the modeled covariance matrix.

Running this way, one can hope to construct the unbiased estimators
accounting for systematic errors through covariance matrix, 
but to be aware of its unbiasness the study of their properties is needed.
In view of lack of the comprehensive information on this scope in 
literature, this paper is devoted to the analysis of the 
statistical properties of such estimators with a particular attention paid
on the control of the bias. Through the paper the CME  
properties are compared with the properties of the 
simplest $\chi^2$ estimator (SCE) as well as if was done 
earlier in Ref.~\cite{Alekhin:1995ij}.

\section{THE SIMPLEST $\chi^2$ ESTIMATOR}

 To illustrate our method of the statistical properties analysis
we start from the analysis of uncorrelated measurements. In this case,
if the data sample $\{y_i\}$ is supposed to be
explicitly described by a theoretical model $t_i=f_i(\theta^0)$,
\begin{equation}
y_i=t_i+\mu_i \sigma_i,
\label{UNCSET}
\end{equation}
where $\mu_i$ are independent random variables, $\sigma_i$ are 
statistical errors,
$i=1\ldots N$, $N$ is the total number of points in the sample. 
We adopt that theoretical model parameter $\theta^0$ is scalar,
the generalization of the 
formula on the case of vector parameter is evident.
If $y_i$ are obtained in the counting experiment with 
the large number of events, $\mu_i$ are Gaussian distributed,
although it is not crucial for our consideration.
As a rule the values of $\sigma_i$ 
given in the experimental publications, are the estimators of the
$y_i$ standard deviations, i.e. are random variables,
but we neglect their fluctuations.
The SCE is based on the minimization of functional
\begin{equation}
\chi^2(\theta)=\sum_{i=1}^{N} \frac{(f_i(\theta)-y_i)^2}{\sigma_i^2}
\label{SIMCHI}
\end{equation}
or, equivalently, solution of the equation
\begin{equation}
\xi(\theta)\equiv\frac{1}{2}\frac{\partial\chi^2}{\partial\theta}=0.
\label{BASEQN}
\end{equation}
The solution $\hat\theta$ is the estimator of
parameter $\theta$, which is the random variable depending on $\{y_i\}$.
To investigate statistical properties of $\hat\theta$ we expand 
the function $\xi(\theta)$
around $\theta^0$ and then apply Legendre inversion to 
obtain the series for $\hat\theta$ (see Ref.~\cite{JAMES} 
for the details of method).
Introducing
\begin{displaymath}
X=\xi(\theta^0),~~~~~~
a=-\left\langle\frac{\partial\xi(\theta^0)}{\partial\theta}\right\rangle ,
\end{displaymath}
\begin{displaymath}
b=\left\langle\frac{\partial^2\xi(\theta^0)}{\partial\theta^2}\right\rangle
,~~~~~Y=\frac{\partial\xi(\theta^0)}{\partial\theta}
-\left\langle\frac{\partial\xi(\theta^0)}{\partial\theta}\right\rangle,
\end{displaymath}
one can obtain
\begin{equation}
\hat\theta-\theta^0=\frac{X}{a}+\frac{X Y}{a^2}+\frac{b X^2}{2a^3}+\ldots
\label{GENBIAS}
\end{equation}
where $<~>$ means averaging over the samples
and the rejected part of the expansion  contains the terms with the higher
powers of $1/a$ and/or $X$ and $Y$.
In this approximation the dispersion of $\hat\theta$ is 
\begin{displaymath}
D(\hat\theta)=\frac{\left\langle X^2\right\rangle}{a^2}
\end{displaymath}
and the bias is 
\begin{displaymath}
B(\hat\theta)=\frac{\left\langle X\right\rangle}{a}+
\frac{\left\langle X Y\right\rangle}{a^2}
+\frac{b\left\langle X^2\right\rangle}{2a^3}.
\end{displaymath}
For the SCE applied to the sample (\ref{UNCSET})
one can easily obtain
\begin{displaymath}
\left\langle X\right\rangle=0,
\end{displaymath}
\begin{displaymath}
\left\langle X^2\right\rangle=-a=\sum_{i=1}^{N} 
\frac{\left[f_i'(\theta_0)\right]^2}{\sigma_i^2},
\end{displaymath}
\begin{equation}
\left\langle X Y\right\rangle=\frac{b}{3}=\sum_{i=1}^{N} 
\frac{f_i'(\theta_0)f_i''(\theta_0)}{\sigma_i^2},
\label{eqn:ab0}
\end{equation}
where $f_i'(\theta)$ is the derivative on $\theta$.
The dispersion and the bias of this estimator are
\begin{equation}
D_0^{\rm U}(\hat\theta)=-\frac{1}{a},~~~~~~
B_0^{\rm U}(\hat\theta)=-\frac{b}{6a^2}.
\label{DISPS}
\end{equation}

If $f_i(\theta)$ are the linear functions of $\theta$ the
series (\ref{GENBIAS}) is truncated 
and equation (\ref{BASEQN}) can be solved exactly. 
One can see that in this case the estimator bias vanishes.
For a non-linear data model the expansion (\ref{GENBIAS})  
contains an infinite number of terms, but
the contributions from the highest terms are
proportional to the powers of $D(\hat\theta$) and/or 
to the central moments of $y_i$ higher than the second.
These contributions are progressively suppressed comparing with the main terms 
if the data statistics rises. Here and through the paper we neglect the 
contribution from the high 
moments of $y_i$. Remind that the same approximation  
is used in deducing of the central limit theorem of statistics.
This approach can 
be also used to justify the analysis of a nonlinear 
data model: The above formula can be applied to 
the data model with a ``weak nonlinearity'', i.e. if its nonlinearity 
is not significant on the scale of the parameter standard deviation.

 Now let the sample to have a common
additive systematic error. In accordance with the Bayesian approach
to the treatment of systematic errors the measured values are given by  
\begin{equation}
y_i=t_i+\mu_i \sigma_i+\lambda s_i,
\label{CADDSET}
\end{equation}
where $s_i$ are systematic shifts for every point and $\lambda$
is the random variable with zero average and unity dispersion\footnote
{Emphasize, that $\lambda$ is not necessary Gaussian distributed.}.
Consider the case of one source of systematic error, generalization
on the many sources case is straightforward.
For the sample (\ref{CADDSET}) we loose statistical independence of
measurements and
with the account of the their correlations the relevant expression
for the dispersion and bias are more complicated
\begin{displaymath}
\left\langle X^2\right\rangle=\sum_{i,j=1}^{N} 
\frac{C_{ij}}{\sigma_i^2 \sigma_j^2}f_i'(\theta_0)f_j'(\theta_0)
=-a+\left[\sum_{i=1}^N \frac{s_i}{\sigma_i^2}
f_i'(\theta_0)\right]^2,
\end{displaymath}
\begin{displaymath}
\left\langle X Y\right\rangle=\sum_{i,j=1}^{N} 
\frac{C_{ij}}{\sigma_i^2 \sigma_j^2}f_i'(\theta_0)f_j''(\theta_0)
\end{displaymath}
\begin{displaymath}
=\frac{b}{3}+\left[\sum_{i=1}^N \frac{s_i}{\sigma_i^2}
f_i'(\theta_0)\right]
\left[\sum_{i=1}^N \frac{s_i}{\sigma_i^2}
f_i''(\theta_0)\right],
\end{displaymath}
where $a$ and $b$ are given by Eqn.~(\ref{eqn:ab0}),
$C_{ij}$ is the covariance matrix for $\{y_i\}$
\begin{equation}
C_{ij} =  s_i s_j+\delta_{ij}\sigma_i\sigma_j,
\label{COVA}
\end{equation}
and $\delta_{ij}$ is Kronecker symbol. Expressions for $a$ and $b$
are the same as for the uncorrelated data case.
In terms of the $N$-component vectors
\begin{displaymath}
\rho_i=\frac{s_i}{\sigma_i},~~~~~
\phi_1^i=\frac{f_i'(\theta_0)}{\sigma_i},~~~~
\phi_2^i=\frac{f_i''(\theta_0)}{\sigma_i}
\end{displaymath}
the dispersion and the bias in this case can be expressed as
\begin{equation}
D_0^{\rm A}(\hat\theta)=\frac{1}{\phi_1^2}\left(1+\rho^2 z^2_1\right),
\label{DISPSA}
\end{equation}
\begin{equation}
B_0^{\rm A}(\hat\theta)=-\frac{\phi_2}
{2\phi_1^3}\left[\Bigl(1+\frac{3}{2}\rho^2 z^2_1\Bigr)z_{12} - 
\rho^2 z_1 z_2\right],
\label{BIASSA}
\end{equation}
where $\rho, \phi_1, \phi_2$ denote the vectors modulus,
$z_1$ is the cosine of angle between $\vec\rho$ and $\vec\phi_1$,
$z_2$ -- between $\vec\rho$ and $\vec\phi_2$,
$z_{12}$ -- between $\vec\phi_1$ and $\vec\phi_2$.
The dispersion of $\hat\theta$ is larger than for uncorrelated data
because now it also accounts for the fluctuations due to 
systematic errors. As to the bias it remains zero for the linear 
model.

 If systematic errors are multiplicative
\begin{equation}
y_i=(t_i+\mu_i \sigma_i)(1+\lambda \eta_i),
\label{CMULSET}
\end{equation}
where $\eta_i$ quantify the systematic errors. If both statistical and 
systematic errors are small comparing with $t_i$
$$
y_i\approx t_i+\mu_i \sigma_i+\lambda \eta_i t_i,
$$
the correlation matrix is
\begin{equation}
C_{ij} =  \eta_i \eta_j t_i t_j+\delta_{ij}\sigma_i\sigma_j,
\label{COVM}
\end{equation}
and the expressions for the bias and dispersion are the
same as for the additive systematics case
after the substitution $s_i \rightarrow \eta_i t_i$.

The Eqn.~(\ref{DISPSA}) can be split into the parts
which correspond to the  
statistical and systematic fluctuations. One can see that when 
vectors $\vec\rho$ and $\vec\phi_1$ are orthogonal
the systematic error on $\hat\theta$ is equal to zero
and the total dispersion is suppressed. 
Such suppression can be illustrated on the example of 
the extraction of asymmetry from 
the data with general offset error.
Let $f_i(\theta)=\theta x_i$ and both statistical and systematic errors are 
constant through the sample: $s_i=s$, $\sigma_i=\sigma$.
Then $\rho_i=s/\sigma$, $\phi_1^i=x_i/\sigma$ and 
$z_1\sim \sum x_i$. If the positive and negative values of $x_i$
compensate each other in 
the measurements, $z_1=0$ and  the systematic error vanishes.
The appropriate data filtration can also be used to suppress the dispersion 
(\ref{DISPSA}). To clarify the mechanism of this suppression let us trace the 
effect of a separate data point on the dispersion value. 
Add to the data set a point with statistical error $\sigma_0$, 
systematic error $s_0$ and the data model $f_0(\theta)$. If the initial 
data set is large and the systematic error is comparable with statistics, i.e.
$$
N \gg 1,~~~~~~~~~~~~~~\rho \gg 1,
$$
\begin{equation}
\phi_1 \gg \frac{f_0'(\theta_0)}{\sigma_0},~~~~~~~~~~  
\rho\phi_1 z_1\gg \frac{s_0}{\sigma_0^2}f_0'(\theta_0), 
\label{eqn:largeset}
\end{equation}
the change of $D_0^{\rm A}(\hat\theta)$ after adding the new point is
\begin{equation}
\Delta D_0^{\rm A}(\hat\theta)\approx\frac{2\rho}{\phi_1^3}
\frac{1}{\sigma_0^2}\left[z_1 s_0f_0'(\theta_0)
-\frac{\rho z_1^2}{\phi_1}\left[f_0'(\theta_0)\right]^2\right].
\label{eqn:deldisp0}
\end{equation}
The second term in brackets is always negative and gives 
the decrease of dispersion 
due to improved statistical precision. At the same time the first term 
can be negative or positive, depending on the signs of $z_1$ and $s_0$.
Its absolute value can be larger than 
the absolute value of the second term and then 
$D_0^{\rm A}(\hat\theta)$ can increase or decrease after adding the new point.
This is manifestation of inconsistency of the SCE applied to the 
correlated data set.
The balance between terms of Eqn.~({\ref{eqn:deldisp0})
 is defined by the distribution 
of $f_i'(\theta_0)/s_i$ and cuts of the tails of this distribution 
can decrease the estimator dispersion.

\section{THE COVARIANCE MATRIX ESTIMATOR}

If systematic error is additive and covariance matrix is known
a priori and is given by (\ref{COVA}) one can use for the parameter 
estimation the following functional minimization 
\begin{equation}
\chi^2(\theta)=\sum_{i,j=1}^{N} (f_i(\theta)-y_i) E_{ij} (f_j(\theta)-y_j),
\label{CORCHI}
\end{equation}
where $E_{ij}$ is the inverted correlation matrix.
This problem can be reduced to the uncorrelated
case using the linear transformation of the vector $\{f_i(\theta)-y_i\}$
and the estimator is linear for the linear data model.
Besides, if statistical and systematics fluctuations obey 
the Gaussian distribution,
this estimator provides minimal dispersion due to the Cramer-Rao
inequality. 

One can easily derive the expressions necessary to calculate the
estimator bias and dispersion
\begin{displaymath}
\left\langle X\right\rangle=0,
\end{displaymath}
\begin{displaymath}
\left\langle X^2\right\rangle=-a=\sum_{i,j=1}^{N}
f_i'(\theta_0)E_{ij}f_j'(\theta_0),
\end{displaymath}
\begin{displaymath}
\left\langle X Y\right\rangle=\frac{b}{3}=\sum_{i,j=1}^{N}
f_i'(\theta_0)E_{ij}f_j''(\theta_0).
\end{displaymath}
Substituting in the above relations the explicit expression for $E_{ij}$
\begin{displaymath}
E_{ij}=\frac{1}{\sigma_i \sigma_j}
\Bigl(\delta_{ij} -
\frac{\rho_i\rho_j}{1+\rho^2}\Bigr)
\end{displaymath}
we obtain the estimator dispersion
\begin{equation}
D_{\rm M}^{\rm A}(\hat\theta)=\frac{1}{\phi_1^2}
\left[1+\frac{\rho^2 z_1^2}{1+\rho^2(1-z_1^2)}\right]
=\frac{1}{\phi_1^2}\xi_{\rm M},
\label{DISPCA}
\end{equation}
where $\xi_{\rm M}$ is the ratio of the total dispersion to the 
pure statistical one.
If $\vec\rho$ and $\vec\phi_1$ are collinear the dispersion of the estimator is
\begin{displaymath}
D^{A,\parallel}_{\rm M}(\hat\theta)=\frac{1+\rho^2}
{\phi_1^2},
\end{displaymath}
which coincide with the SCE dispersion (\ref{DISPSA}).
One can see that if $\vec\rho$ and $\vec\phi_1$ are not collinear
the SCE dispersion (\ref{DISPSA})
is always larger than the CME dispersion (\ref{DISPCA}).
This can be readily explained qualitatively. 
For SCE the fitted curve tightly follows the 
data points and, if these points are shifted due to the systematic errors
fluctuations, the parameter gains appropriate systematic errors. 
At the same time, since for the CME the information on the data 
correlations is explicitly included in $\chi^2$, the correlated  
fluctuation of the data due to systematic shift does not necessary leads to 
the fitted curve shift and the parameter deviation gets smaller than 
for SCE. 
The exclusion occurs if $z_1=0$, when $\vec\rho$ and $\vec\phi_1$ are collinear 
and the systematic shift can be perfectly 
compensated by the change of parameter.
If these vectors are orthogonal the CME dispersions is 
\begin{displaymath}
D^{A,\perp}_{\rm M}(\hat\theta)=\frac{1}{\phi_1^2}
\end{displaymath}
i.e. it is just the same as the dispersion of SCE  
applied to the data set without correlations (\ref{DISPS}).
Qualitatively it corresponds to the measurements scheme when
systematic shift for the different points compensate each other,
e.g. as in the example considered at the end of Sec.~1.

For the modern experiments systematic errors are often of the same order 
as statistical ones and if $N\gg 1$ then $\rho\gg 1$.
In this limit and if $\vec\rho$ and $\vec\phi_1$ are not collinear
\begin{equation}
D_{\rm M}^{\rm A}(\hat\theta)\approx\frac{1}{\phi_1^2(1-z^2_1)}
\label{eqn:dispmr}
\end{equation}
and
\begin{equation}
D_0^{\rm A}(\hat\theta)\approx\frac{\rho^2 z^2_1}{\phi_1^2}.
\label{eqn:disp0r}
\end{equation}
One can see that in the second case 
the estimator standard deviation rises linearly with
the increase of the systematics, whereas the CME dispersion saturates. 
This difference can be illustrated on the numerical example
inspired by the elastic proton-proton scattering. Let us choose
$$
f_i=U\exp^{(-V x_i)},~~~~x_i=0.1 i,
$$
where $U=100, V=10, i=1\ldots 9$.
Generating 100 data sets (\ref{CADDSET}) with these $f_i$ and
\begin{equation}
\sigma_i=0.01\sqrt{\frac{U}{f_i}},~~~~s_i=\frac{\kappa}{x_i}
\label{eqn:testset}
\end{equation}
we minimized functionals (\ref{SIMCHI}) and (\ref{CORCHI}) varying
$U$ and $V$ to obtain their estimators $\hat U$ and $\hat V$.
The values of $(\hat U-U)^2$ and $(\hat V-V)^2$
for all of the generated data sets were averaged to obtain the
estimators dispersions.
The results on the standard deviation of $\hat U$ for different values of
$\kappa$ are given in 
Fig.~\ref{fig:disp} (the picture for $\hat V$ is similar).
One can see that at large $\kappa$ the CME and the SCE standard deviations   
differ by factor of 3.

The example of dispersion suppression observed in
the analysis of real experimental 
data can be found in Ref.~\cite{Alekhin:1995dz}.
In this paper we performed the 
leading order QCD fit to the inclusive deep inelastic scattering data 
of Refs.~\cite{Benvenuti:1989rh,Benvenuti:1990fm}
obtained by the BCDMS collaboration 
in order to determine the parton distribution functions and the 
strong coupling constant value $\alpha_{\rm s}$. The two different estimators
were used and the different estimates were obtained. For the 
SCE the standard deviation of $\alpha_{\rm s}(M_{\rm Z})$
is 0.015, while for the CME it is 0.007.
The difference in the gluon distribution bounds for these estimators
can is given in Fig.~\ref{fig:bcdms}. One can see that the standard deviation 
of the gluon distribution for the CME is also about 
a half of the SCE standard deviation.

If $z_1\ne 1$, the change of CME dispersion
after adding a new point to the large sample as defined by  
Eqn.~(\ref{eqn:largeset}) is 
\begin{displaymath}
\Delta D_{\rm M}^{\rm A}(\hat\theta)\approx-\frac{1}{\phi_1^4(1-z_1^2)^2}
\frac{1}{\sigma_0^2}
\left[f_0'(\theta_0)
-\frac{\phi_1 z_1}{\rho}s_0\right]^2.
\end{displaymath}
This change is always negative that proves the CME
consistency. Remind, that this is not necessary for the 
SCE (see Sec.~1). The same conclusion can be drawn 
from the comparison of Eqns.~(\ref{eqn:dispmr})
and (\ref{eqn:disp0r}). Indeed, the CME dispersion falls with 
increase of statistical significance of the data set (i.e. decrease of 
$\sigma$ or rise of $N$) while the SCE dispersion does not.
Note, that due to consistency of the CME 
the filtration procedure described in Sec.~1 
is not meaningful for it.

\begin{figure}[t]
\centerline{\psfig{figure=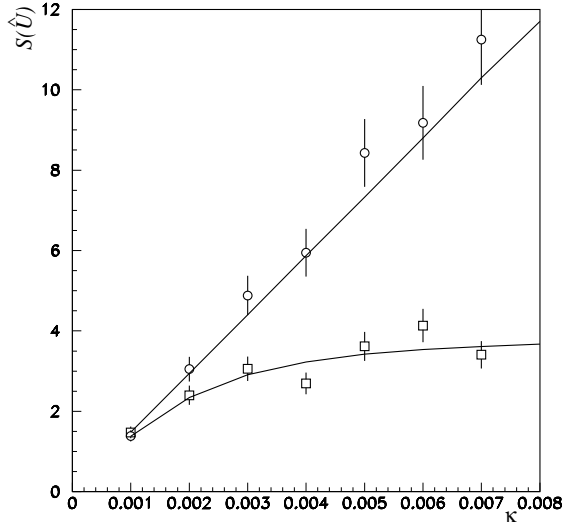,height=7cm}}
\caption{The standard deviations of SCE (circles) and CME (squares)
for $\hat U$ at different scales of systematic errors $\kappa$.
The lines correspond to the calculation performed with 
the two-dimensional generalization of
Eqns.~(9,16).}
\label{fig:disp}
\end{figure}

The CME bias is
\begin{displaymath}
B_{\rm M}^{\rm A}(\hat\theta)=-\frac{\phi_1\phi_2}{2}
\left[D_{\rm M}^{\rm A}(\hat\theta)\right]^2\left(z_{12}
-\frac{\rho^2}{1+\rho^2}z_1 z_2\right),
\end{displaymath}
which vanishes for the linear data model and 
saturates in the limit of $\rho\gg 1$ contrary to the SCE. 
In the numerical example (\ref{eqn:testset}) at $\kappa=0.007$
the CME bias is 0.07, whereas the SCE bias is 0.13.

\begin{figure}[t]
\centerline{\psfig{figure=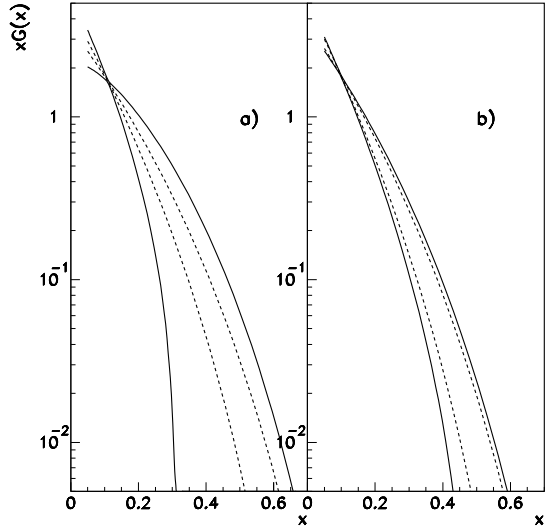,height=7cm}}
\caption{Bounds of gluon distribution obtained from the 
LO QCD fit to BCDMS data 
with different estimators (the SCE: a; the CME: b).
Full lines correspond to the total experimental errors, dashed ones -- to
the statistical only.}
\label{fig:bcdms}
\end{figure}

For the multiplicative systematic errors
the covariance matrix in unknown a priori and one is to calculate 
it using the parameter estimator. Proceeding this
way in the minimization of the functional (\ref{CORCHI}) we get 
\begin{equation}
a=-\sum_{i,j=1}^{N}f_i'(\theta^0)E_{ij}f_j'(\theta^0)-
\frac{1}{2}\sum_{i,j=1}^{N}E_{ij}''C_{ij}.
\label{eqn:dispm}
\end{equation}
The difference with corresponding expression for the case 
of additive systematic errors
is in the second term of Eqn.~(\ref{eqn:dispm}).
For the linear data model this term is
$$
a^{(2)}=\frac{1}{2}\sum_{i,j=1}^{N}E_{ij}''C_{ij}=
\frac{\phi_3^2}{2(1+\rho^2)^2}
\left[\rho^4(z^2_3-1)-3\rho^2 z^2_{3}+1\right],
$$
where
\begin{displaymath}
\phi_3^i=\rho_i'=\frac{\rho_i}{f_i}f_i'(\theta^0)=\eta_i\phi_1^i,
\end{displaymath}
$\phi_3$ is modulus of $\vec\phi_3$
and $z_{3}$ is the cosine of the angle between $\vec\phi_3$ and $\vec\rho$.
The ratio of the second term of Eqn.~(\ref{eqn:dispm}) to the first term
$a^{(1)}=\sum f_i'(\theta^0)E_{ij}f_j'(\theta^0)$ is 
\begin{equation}
\frac{a^{(2)}}{a^{(1)}}=
\frac{\phi_3^2}{\phi_1^2}\cdot
\frac{\rho^4(z^2_3-1)-3\rho^2 z^2_{3}+1}
{(1+\rho^2)^2}\xi_{\rm M}.
\label{eqn:dispmult}
\end{equation}
If $\xi_{\rm M}\sim O(1)$ (that is valid for most real cases),
$a^{(2)}\sim O(\eta^2)a^{(1)}$ for all values of $\rho$, i.e. it
is suppressed comparing with the first term for small $\eta$.
Neglecting as elsewhere the third and fourth central moments of $\{y_i\}$, 
one can obtain that $<X^2>\approx -a$ and 
the estimator dispersion for multiplicative 
systematic errors $D_{\rm M}^{\rm M}\approx D_{\rm M}^{\rm A}$ .

In the case of multiplicative systematics errors Eqn.~(\ref{BASEQN})
is nonlinear even for the linear data model.
As a consequence, the expressions responsible for the bias
$$
\left\langle X \right\rangle=
\frac{1}{2}\sum_{i,j=1}^{N} E_{ij}' C_{ij},
$$
$$
b=3 \sum_{i,j=1}^{N}
f_i'(\theta^0)E_{ij}f_j''(\theta^0)+
3 \sum_{i,j=1}^{N}f_i'(\theta^0) E_{ij}'f_j'(\theta^0)+
\frac{1}{2}\sum_{i,j=1}^{N}E_{ij}'''C_{ij},
$$
\begin{equation}
\left\langle X Y\right\rangle=\sum_{i,j=1}^{N}
f_i'(\theta^0)E_{ij}f_j''(\theta^0)+
2 \sum_{i,j=1}^{N}
f_i'(\theta^0) E_{ij}' f_j'(\theta^0)-
\frac{1}{4}\sum_{i,j=1}^{N}E_{ij}''C_{ij}\sum_{i,j=1}^{N} E_{ij}' C_{ij},
\label{eqn:biasm}
\end{equation}
do not vanish even if $f_i''({\theta})$ is equal to zero.
Meanwhile the bias due to the estimator nonlinearity is small comparing with 
the estimator standard deviation. Since 
$1/D_{\rm M}^{\rm M}\approx <X^2>\approx -a$ the bias of estimator with 
multiplicative systematic errors is
\begin{equation}
B_{\rm M}^{\rm M}(\hat\theta)\approx\sqrt{D_{\rm M}^{\rm M}(\hat\theta)}
\left[\frac{\left\langle X\right\rangle}{\sqrt{-a}}+
\frac{\left\langle X Y\right\rangle-b/2}{(-a)^{3/2}}\right].
\label{eqn:biasmult}
\end{equation}
The first term in the brackets of Eqn.~(\ref{eqn:biasmult}) is
\begin{equation}
\frac{\left\langle X\right\rangle}{\sqrt{-a}}\approx
-\frac{\phi_3}{\phi_1}
\frac{\rho z_3}{1+\rho^2}\sqrt{\xi_{\rm M}}\sim O(\eta\sqrt{\xi_{\rm M}}).
\label{eqn:biasx}
\end{equation}
The contribution to the second term in brackets of Eqn.~(\ref{eqn:biasmult})
from $\sum f_i'(\theta^0) E_{ij}'f_j'(\theta^0)$ is proportional to 
$$  
\frac{\phi_3}{\phi_1}\frac{\rho z_1}{(1+\rho^2)} 
\left(\frac{\rho^2}{1+\rho^2}z_1 z_3-z_{13}\right)\xi_{\rm M}^{3/2}
$$
and hence it is $\sim O(\eta\xi_{\rm M}^{3/2})$. As one can conclude from 
Eqns.~(\ref{eqn:dispmult},\ref{eqn:biasx}) the contribution 
to the same term from 
$\sum E_{ij}''C_{ij}\cdot\sum E_{ij}' C_{ij}$ is $O(\eta^3\xi_{\rm M}^{3/2})$. 
And finally since 
$$
\frac{1}{2}
\sum_{i,j=1}^N E_{ij}'''C_{ij}=\frac{\rho z_1\phi_3^3}{(1+\rho^3)^2}
\left[\rho^4(z_3^2-1)+\rho^2(1-3z_3^2)+2\right]
$$
the contribution to Eqn.~(\ref{eqn:biasmult}) coming from this term  
is $O(\eta^3\xi_{\rm M}^{3/2})$. 
In summary, for the linear data model
the estimator bias is a sum of terms 
$O(\eta^{p}\xi_{M}^{q})D^{\rm M}_{\rm M}$ 
with $p\ge 1$ and $q\le 3/2$.  
Besides, at small $\rho$ all the four contributions to the bias
which survive for the linear data model are 
$\sim \rho$ while at large $\rho$
they are  $\sim1/\rho$. Summarizing, one can conclude that 
the estimator is negligible excluding the extreme cases with very large 
$\xi_{\rm M}$.

The explicit estimate of the bias can be obtained from the 
Eqns.~(\ref{eqn:biasm},\ref{eqn:biasmult}). 
Meanwhile it requires rather lengthy calculations and more 
simple tool for the bias evaluation is admirable.
A convenient way for this is to trace the net residual
\begin{displaymath}
R=-\frac{1}{N}\sum_{i=1}^{N}\frac{f_i(\hat\theta)-y_i}
{\sqrt{\sigma_i^2+s_i^2}}.
\end{displaymath}
Expanding $f_i(\theta)$ near $\theta_0$ 
and keeping only the first term in Eqn.~(\ref{GENBIAS})
one obtains for the sample (\ref{CADDSET})
$$
R\approx -\frac{1}{N}
\sum_{i=1}^{N}\frac{\mu_i+\lambda \rho_i}{\sqrt{1+\rho_i^2}}
+(\hat\theta-\theta_0)
\frac{1}{N}\sum_{i=1}^{N}\frac{\phi_1^i}{\sqrt{1+\rho_i^2}}.
$$
If the estimator is unbiased, the value of $R$ averaged over 
the samples is equal to zero. Nevertheless the particular values of $R$ 
may be not equal to zero due to fluctuations. 
For the limited $\xi_{\rm M}$ the dispersion of $R$ is
\begin{equation}
D(R)=\frac{1}{N^2}\sum_{i,j=1}^{N}\frac{\delta _{ij}+\rho_i\rho_j}
{\sqrt{1+\rho_i^2}\sqrt{1+\rho_j^2}}+O(1/N).
\label{eqn:Rdisp}
\end{equation}
If the analyzed data come from a single experiment 
with dominating systematics (i.e with $ \rho > 1$) then 
$D(R)\sim 1$. In particular for the BCDMS data 
of Refs.~\cite{Benvenuti:1989rh,Benvenuti:1990fm} $D(R)\approx 0.7$.
For $N_{exp}$ independent experiments involved in the analysis
$D(R)\sim 1/N_{exp}$. Comparing the net residual $R$
with this value allows to get a guess about the estimator bias.
More definite conclusion 
can be drawn after the comparison of $R$
with its dispersion calculated using Eqn.~(\ref{eqn:Rdisp}).

\section{PLANNING OF THE COUNTING EXPERIMENTS} 

In a particular case when the differential cross section on the
variable $x$ is measured, the predicted average 
number of events in the $i-$th bin of is 
$$
\left\langle N_i\right\rangle =Lf_i\Delta x_i\beta_i,
$$
where $L$ is the integral experiment luminosity,
$\beta_i$ is the registration efficiency, and $\Delta x_i$ is 
the bin width. Neglecting the fluctuations of $N_i$
the statistical error on the $i-$th measurement is 
$$
\sigma_i=\frac{\sqrt{\left\langle N_i\right\rangle}}{L\Delta x_i\beta_i}
$$
and 
$$
\frac{1}{\sigma_i^2}=\frac{L\beta_i}{f_i}\Delta x_i.
$$
The scalar product of the vectors $\vec \rho$ and $\vec \phi$
is 
$$ 
\left(\vec\rho \cdot \vec\phi\right)=L\sum_{i=1}^{N}\frac{f_i's_i}
{f_i}\beta_i\Delta x_i
$$
and
$$ 
\phi^2=L\sum_{i=1}^{N}\frac{\left[f_i'\right]^2}
{f_i}\beta_i\Delta x_i,~~~~~~
\rho^2=L\sum_{i=1}^{N}\frac{\left[s_i\right]^2}
{f_i}\beta_i\Delta x_i.
$$
For the dense measurements these scalars can be  
reduced to the integrals over the measurements region $\Omega$: 
$$ 
\left(\vec\rho \cdot \vec\phi\right)=L\int_{\Omega} f'(x)s(x) d\tilde{x}
$$
and
$$ 
\phi^2=L\int_{\Omega}\left[f'(x)\right]^2 d\tilde x,~~~~~~
\rho^2=L\int_{\Omega}\left[s(x)\right]^2 d\tilde x,
$$
where $d\tilde{x}=\beta(x)/f(x) dx$.
The latter expressions can be used in 
the equations for the estimators dispersions\footnote{As a result 
one obtains the Fisher's information for the 
correlated data case.}.
This approach is convenient for the future experiment optimization 
since it allows for to analyze integrated expression  
in order to search for the optimal region of measurements. 
For the simple functions $f(x)$, $\beta(x)$, and $s(x)$ such 
analysis sure can be performed analytically.

\section{CONCLUSION}

In conclusion, the CME is a convenient tool 
for the analysis of the data sets with the account of correlations due to 
systematic errors. The CME is consistent 
for the realistic cases (when systematic errors on the fitted parameters are 
not extremely large comparing with the statistical ones)
and its dispersion is always smaller, than 
the dispersion of the $\chi^2$ estimator without account of correlations.
The estimator bias is negligible for the realistic 
cases if the covariance matrix is calculated during the fit iteratively 
using the parameter estimator itself. Analytical formula for the covariance 
matrix inversion allows to perform fast and precise calculations 
even for very large data sets. The latter 
is especially important in view of numerical instabilities occurring 
in the fits to precise data in the case of large correlation between 
the fitted parameters (see in this connection 
Ref.~\cite{Alekhin:1994}).

A particular attention should be paid on 
the connection between the estimator dispersion 
and the confidence interval. For a known distribution 
of the estimator the confidence interval 
can be easily calculated
(e.g. it is well known that for the Gaussian distribution 
one standard deviation corresponds to the 67\% confidence level).
Unfortunately due to the possible non-Gaussian nature of the systematic errors
one cannot prove that an estimator accounting for systematics
is Gaussian distributed.
However for the large number of systematic errors of comparable scale
the estimator should obey Gaussian distribution just to the central 
limit theorem of statistics. Otherwise the robust estimates of the 
confidence intervals, e.g. Chebyshev's inequality, should be used.

{\bf Acknowledgments}

I am indebted to S.Keller for valuable discussions and comments. 
The work was supported by RFBR grant 00-02-17432.

\end{document}